\documentclass{PoS}
\usepackage{amsmath}
\usepackage{amssymb}
\usepackage{graphicx}
\usepackage{cite}
\usepackage{wrapfig}
\usepackage{overpic}
\newcommand{\kt}{{\mathbf k}}
\newcommand{\qt}{{\mathbf q}} 
\newcommand{\pt}{{\mathbf p}}

\definecolor{red}{rgb}{1,0,0}
\def\lesssim{\ \hbox{\raise 2pt \hbox{$<$} \kern -13pt
                     \lower 3pt \hbox{$\sim$}}\ }
\def\greatersim{\ \hbox{\raise 2pt \hbox{$>$} \kern -13pt
                     \lower 3pt \hbox{$\sim$}}\ }

\def\cascade{{\sc Cascade}}

\def\powheg{{\sc Powheg}}
\def\mcatnlo{{\sc Mc@nlo}}

\newcommand{\mincas}{{\sf MINCAS}}
\input epsf.tex
\def\desepsf(#1 width #2){\epsfxsize=#2 \epsfbox{#1}}
\def\kt{\ensuremath{k_t}}

\def\pt{\ensuremath{p_t}}
\def\qt{\ensuremath{q_t}}

\newcommand{\as}{\ensuremath{\alpha_\mathrm{s}}}

\def\katie{{\sc KaTie}}

\title{QCD theory overview}

\ShortTitle{QCD}

\author{\speaker{Krzysztof Kutak}\\
        Institute of Nuclear Physics Polish Academy of Sciences,\\
        Radzikowskiego 152, 31-342, Krakow\\
        E-mail: \email{krzysztof.kutak@ifj.edu.pl}}

\abstract{In this review I am going to present biased by personal  experience review of QCD. The review covers topics 
like Improved Transverse Momentum Dependent Factorization, $k_T$ dependent splitting functions, Transversal Momentum Dependent parton shower, non-Gaussian broadening of jet traversing quark gluon plasma.}

\FullConference{Corfu Summer Institute 2019 "School and Workshops on Elementary Particle Physics and Gravity" (CORFU2019)\\
		31 August - 25 September 2019\\
		Corfù, Greece}

\begin{document}
\section{Factorization for forward dijet production}
High energy collisions of protons and heavy nuclei at the Large Hadron Collider (LHC) provide a unique tool to probe
dense systems of quarks and gluons.
In particularly interesting are processes where jets or particles are produced in the forward direction with respect to the incoming proton. Kinematically, such final  states have large rapidities and therefore they trigger events in which the partons from the nucleus carry rather small longitudinal momentum fraction $x$.
This kinematic setup is perfectly suited to investigate the phenomenon of gluon saturation, which is expected to occur at some value of $x$ to prevent violation of the unitarity bound (for a review of this subject see Ref. \cite{Albacete:2014fwa}). The behaviour of dense systems of partons when $x$ becomes small is predicted by  Quantum Chromodynamics (QCD) and leads to non-linear evolution equations known as B-JIMWLK equations (for review see \cite{Kovchegov:2012mbw,Gelis:2010nm}), which can be derived within the Color Glass Condensate (CGC) theory.
In CGC, the calculation of forward jet production in dense-dilute collisions relies on the hybrid factorization \cite{Dumitru:2005gt,Marquet:2007vb,Deak:2009xt,Deak:2010gk,Deak:2011ga}, where the large-$x$ projectile is described by the collinear PDFs, while the dense target according to theoretical results is described with linear or nonlinear equations depending whether the $x$ of target gluons is moderate or small. The  description of multi-jet production is rather complicated even in this simplified framework~\cite{Marquet:2007vb}. A novel approach to such processes was initiated in Ref.~\cite{Dominguez:2011wm} for dijets in the back-to-back correlation regime and in Ref.~\cite{Kotko:2015ura} for a more general kinematical configuration. The latter is known as the small-$x$ Improved Transverse Momentum Dependent (ITMD) factorization. The ITMD formula accounts for: 
\begin{itemize}
\item complete kinematics of the scattering process with off-shell gluons, 
\item gauge invariant definitions of the TMD gluon densities, 
\item gauge invariant expressions for the off-shell hard matrix elements, 
\item it also recovers the high energy factorization (aka $k_T$-factorization)~\cite{Catani:1990eg,Collins:1991ty,Deak:2009xt} in the limit of large off-shellness of the initial-state gluon from the nucleus. 
\end{itemize}
\begin{figure}[t!]
  \begin{center}
    \includegraphics[width=0.99\textwidth]{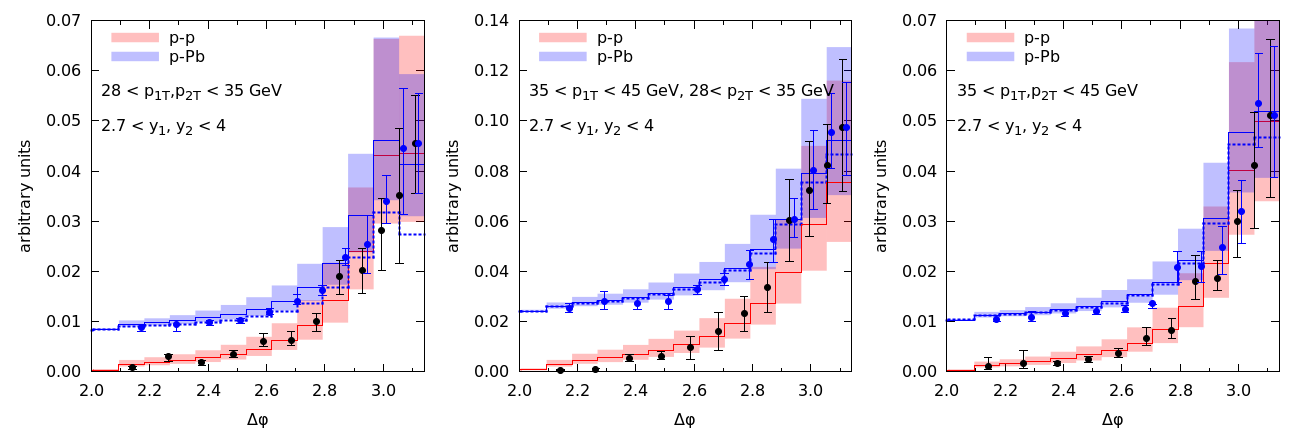}
  \end{center}
  \caption{
    Broadening of  azimuthal decorrelations in p-Pb collisions vs p-p collisions
    for different sets of cuts imposed on the jets' transverse momenta.
    The plots show  normalized cross sections as functions of the azimuthal distance between the two leading jets, $\Delta\phi$. 
    The points show the experimental data \cite{Aaboud:2019oop} for p-p and p-Pb, where the p-Pb data were shifted by a pedestal, so that the values in the bin $\Delta\phi\sim \pi$ are the same.
    Theoretical calculations are represented by the histograms with uncertainty bands coming from varying the scale by factors 1/2 and 2.
    The blue solid and blue dotted lines corresponds to the choices of d=0.5 and d=0.75, respectively.
  }
  \label{fig:broadening}
\end{figure}
Recently, the ITMD  factorization has been proved ~\cite{Altinoluk:2019fui}. Steps in further extension of the formalism to three and more jets were undertaken in Ref.~\cite{Bury:2018kvg,Bury:2020ndc} and in \cite{Altinoluk:2018byz} in the correlation limit.
For some of phenomenological application of the formalizm see \cite{vanHameren:2016ftb, Kotko:2017oxg, Albacete:2018ruq}.
While the original ITMD formula, as well as the works studying the jet correlation limit within CGC, include gluon saturation effects, they do not account for all contributions proportional to logarithms of the hard scale set by the  large transverse momenta of jets -- the so-called Sudakov logarithms. It has been shown in Refs.~\cite{vanHameren:2014ala,vanHameren:2015uia} that inclusion of Sudakov logarithms is necessary in order to describe the LHC jet data at small $x$ but yet before the saturation regime. In the low $x$ domain, the resummation leading to the Sudakov logarithms has been developed in ~\cite{Mueller:2013wwa,Mueller:2012uf,Sun:2014gfa,Mueller:2015ael,Zhou:2016tfe,Xiao:2017yya,Zheng:2019zul} see also \cite{Kutak:2014wga}. In the paper \cite{vanHameren:2019ysa}, it has been shown for the first time, that the interplay of saturation effects and the resummation of the Sudakov logarithms is essential to describe the small-$x$ forward-forward dijet data.
The process under consideration is the inclusive dijet production
\begin{equation}
  \mathrm{p} \left(P_{\mathrm{p}}\right) + \mathrm{A} \left(P_{\mathrm{A}}\right) \to j_1 (p_1) + j_2 (p_2)+ X\ ,
\end{equation}
where $A$ can be either the lead nucleus, as in p-Pb scattering, or a proton, as in p-p scattering. 
To describe the above process,
the hybrid approach has been used where one assumes that the proton $p$ is a dilute
projectile, whose partons are collinear to the beam and carry momenta $p=x_{\mathrm{p}} P_{\mathrm{p}}$.  
The nucleus $A$ is probed at a dense state. 
The jets $j_1$ and $j_2$ originate from hard partons produced in a collision of the collinear parton $a$
with a gluon belonging to the dense system $A$. This gluon is off-shell, with momentum 
$k=x_{\mathrm{A}} P_{\mathrm{A}} + k_T$ and $k^2=-|\vec{k}_T|^2$.
The ITMD factorization formula for the production of two jets with momenta $p_1$ and $p_2$, and rapidities $y_1$ and $y_2$, reads
\begin{equation}
\frac{d\sigma^{\mathrm{pA}\rightarrow j_1j_2+X}}{d^{2}q_{T}d^{2}k_{T}dy_{1}dy_{2}}
=
\sum_{a,c,d} x_{\mathrm{p}} f_{a/\mathrm{p}}\left(x_{\mathrm{p}},\mu\right) 
\sum_{i=1}^{2}\mathcal{K}_{ag^*\to cd}^{\left(i\right)}\left(q_T,k_T;\mu\right)
\Phi_{ag\rightarrow cd}^{\left(i\right)}\left(x_{\mathrm{A}},k_T,\mu\right)\,,
\label{eq:itmd}
\end{equation}   
The distributions $f_{{a/\mathrm{p}}}$ are the collinear PDFs corresponding to the large-$x$ gluons and quarks in the projectile. 
The functions $\mathcal{K}_{^{ag^*\to cd}}^{_{(i)}}$ are the hard matrix elements constructed from gauge-invariant off-shell amplitudes \cite{vanHameren:2012uj,vanHameren:2012if,Kotko:2014aba,Antonov:2004hh}. 
The quantities $\Phi_{^{ag\rightarrow cd}}^{_{(i)}}$ are the TMD gluon distributions introduced in Ref.~\cite{Kotko:2015ura} and parametrize a dense state of the nucleus or the proton in terms of small-$x$ gluons, see Ref.~\cite{Petreska:2018cbf} for an overview.
The phase space is parametrized in terms of the final state rapidities of jets $y_1,y_2$, as well as the momenta $\vec{k}_T = \vec{p}_{1T}+\vec{p}_{2T}$. The azimuthal angle between the final state partons is
$\Delta\phi$.
The collinear PDFs, hard matrix elements, and the TMD gluon distributions all depend on the factorization/renormalization scale $\mu$. At leading order, the matrix elements depend on $\mu$ only through the strong coupling constant. The collinear PDFs obey the DGLAP evolution when the scale $\mu$ changes. The evolution of the  TMD gluon distributions is more involved. Typically, in saturation physics, one keeps $\mu$ fixed at some scale of the order of the saturation scale $Q_s$, and performs the evolution in $x$ using the B-JIMWLK\cite{Kovner:1999bj,Nikolaev:2003zf,Iancu:2000hn,Balitsky:1995ub} equation  or its mean field approximation -- the BK equation \cite{Kovchegov:1999yj}.
To apply the ITMD formula one needs to construct the ITMD densities. The Transversal Moentum Dependent gluon densities entering the formula (\ref{eq:itmd}) for lead and for the proton are constructed from distributions given by the KS gluon density~\cite{Kutak:2012rf} and obtained in~\cite{vanHameren:2016ftb}.
The comparison of the obtained cross section to data was possible since
the ATLAS collaboration studied azimuthal correlations of dijets in proton-lead (p-Pb) and proton-proton (p-p) collisions at the center-of-mass energy $\sqrt{s_{NN}}=5.02\,\mathrm{TeV}$ covering the forward rapidity region between $2.7-4.0$ units \cite{Aaboud:2019oop}. The measurement indicates sizable nuclear effects at small values of $x$. 
Fig.~\ref{fig:broadening}  shows normalized cross sections as functions of $\Delta\phi$ in p-p  and p-Pb collisions. 
The three panels correspond to three different cuts on the transverse momenta of the two leading jets.
The points with error bars represent experimental data from Ref.~\cite{Aaboud:2019oop}. 
The main results for p-Pb collisions  are represented by blue solid lines in Fig.~\ref{fig:broadening}. 
The visible broadening comes from the interplay of the non-linear evolution of the initial state and the Sudakov resummation.
One should emphasize that this is a highly non-trivial consequence of the two components present in our theoretical framework: gluon saturation at low $x$ and Sudakov resummation.
\section{Transverse Momentum Dependent splitting functions}
\begin{wrapfigure}{r}{0.4\textwidth}
\centering
\vspace{-0.5cm}
\begin{overpic}[width=0.29\textwidth]{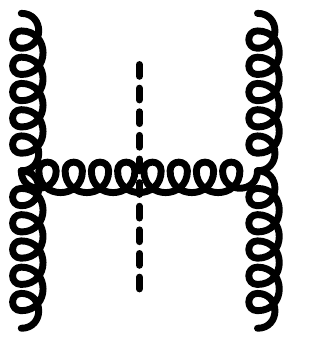}
 \put(-7,20){{\large $k$}}
 \put(-12,83){{\large $z,q$}}
 \put(20,37){{\large $p'$}}
 \put(18,58){{\large $1-z$}}
\end{overpic}
\vspace{-0.5cm}
\caption{Diagram contributing to the real $\tilde{P}_{gg}$ splitting function at leading order.
}
\label{fig:pgg}
\end{wrapfigure}
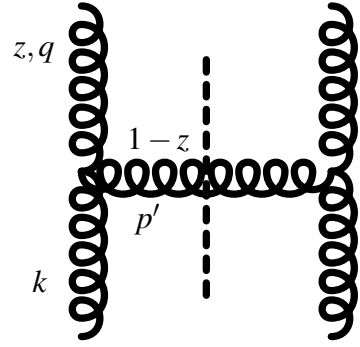
In this section I am going to review an effort to generalize the 
Balitsky - Fadin - Kuraev - Lipatov (BFKL)~\cite{Fadin:1975cb,Kuraev:1976ge,Kuraev:1977fs,Balitsky:1978ic}
evolution equation to larger values of $x$ and to match it to Dokshitzer - Gribov - Lipatov - Altarelli - Parisi (DGLAP) evolution equation.
The BFKL equation is based on the so called $k_T$ (or high-energy)
factorization~\cite{Catani:1990eg} which is strictly speaking valid in the
high energy limit, $s \gg Q^2$ where $Q^2$ is a hard momentum scale e.g virtuality of photon.
If one naively extrapolates the formalism to intermediate or large
$x$, the framework is naturally confronted with a series of
problems and short-comings, e.g. contributions of quarks to
the evolution arise as a pure next-to-leading order (NLO) effect and
elementary vertices violate energy conservation {\it i.e.}
conservation of the longitudinal momentum fraction.
One can account for such effects by including a resummation of terms
which restore subleading, but numerically relevant, pieces of the
DGLAP
\cite{Gribov:1972ri,Altarelli:1977zs,Dokshitzer:1977sg} splitting functions
\cite{Hentschinski:2012kr},\cite{Hentschinski:2013id},\cite{Kwiecinski:1997ee}.
Even though these resummations have
been successful in stabilizing low $x$ evolution into the region of
intermediate $x \sim 10^{-2}$, extrapolations to larger values of $x$ are still prohibited \footnote{{For other approaches to this problem we refer the Reeder to \cite{Deak:2020zay,Ciafaloni:2007gf,Ciafaloni:2006yk} }}. 
To arrive at a framework which avoids the need to account for
kinematic effects through the calculation of higher order corrections,
it is therefore necessary to devise a scheme which accounts for
both DGLAP (conservation of longitudinal momentum) and
BFKL (conservation of transverse momentum) kinematics. There, the low $x$ resummed DGLAP
splitting functions have been constructed following the definition of
DGLAP splittings by Curci-Furmanski-Petronzio (CFP)~\cite{Curci:1980uw}.
The authors of~\cite{Catani:1994sq} were able to define a TMD
gluon-to-quark splitting function $P_{qg}$, both
exact in transverse momentum and longitudinal momentum fraction.
Following observation of~\cite{Hautmann:2012sh}   and derivation of $P_{qg}(z,\qt,\pt^{\prime})$ which made the splitting function transversal momentum dependent the generalization of this scheme to other transition kernels involving quarks $P_{gq}$ and $ P_{qq}$ have been achieved~\cite{Gituliar:2015agu}. 
The computation of the gluon-to-gluon splitting $P_{gg}$ required
a further modification of the formalism used in~\cite{Catani:1994sq,Gituliar:2015agu}
which was recently achieved in~\cite{Hentschinski:2017ayz}.
In order to calculate ${P}_{gg}$ splitting the formalism
of~\cite{Catani:1994sq,Gituliar:2015agu} had to be extended the gluon case by generalizing definition of projector operators and defining appropriate generalized 3-gluon vertex that is gauge invariant in
the presence of the off-shell momentum $k$. More details on the exact procedure can be found in~\cite{Hentschinski:2017ayz}.
The splitting functions reduce both to the conventional gluon-to-gluon DGLAP
splitting in the collinear limit as well as to the LO BFKL
kernel in the low $x$/high energy limit; moreover the CCFM
gluon-to-gluon splitting function is re-obtained in the limit where
the transverse momentum of the emitted gluon vanishes, {\it i.e.} if
the emitted gluon is soft.  The derivation of this result is based on
the Curci-Furmanski-Petronzio formalism for the calculation of DGLAP
splitting functions in axial gauges. 
The next step in completing the calculation of TMD splitting functions
is the determination of the still missing virtual corrections.
With the complete set of splitting functions at hand, it will be finally
possible to formulate an evolution equation for the unintegrated (TMD)
parton distribution functions including both gluons and quarks.
\section{Monte Carlo with Transverse Momentum Dependent shower}
Here I am going to review results obtained in \cite{Bury:2017jxo}. For slightly
different approach but with the same basic principles see \cite{Hautmann:2017fcj,Martinez:2018jxt,Hautmann:2019biw}. While calculations in fixed order perturbation theory in Quantum Chromodynamics (QCD) even at next-to-leading (or even next-to-next-to-leading) order expansion in the strong coupling $\alpha_s$ are often not sufficient, the predictions can be improved when parton showers  are included to simulate even higher order corrections, as done for example with the \powheg\ ~\cite{Alioli:2010xa,Frixione:2007vw} or \mcatnlo\ ~\cite{Frixione:2006gn} methods. However, when supplementing a calculation of collinear initial partons with parton showers, the kinematics of the hard process are changed due to the transverse momentum generated in the initial state shower. This effect can be significant even at large transverse momenta, as has been discussed and shown explicitly in \cite{Hautmann:2012dw}.
With the development of transverse momentum dependent (TMD) parton distributions, this problem can be overcome, since the transverse momentum of the initial partons can be obtained from the TMD parton distributions.  The great advantage of using TMD parton densities is that a  parton shower will not change the kinematics of the matrix element process, in contrast to the conventional approach of collinear hard process calculations supplemented with parton showers, and that the main parameters of the TMD parton shower are fixed with the determination of the TMD.
Already some time ago a TMD parton shower has been developed for the case of initial state gluons within the frame of the CCFM evolution equation \cite{Catani:1989yc,Ciafaloni:1987ur} and implemented in the \cascade\ package \cite{Jung:2010si,Jung:2001hx,Jung:2000hk}.  However, TMD parton densities defined over a large range in $x$, $k_T$  and scale $\mu$ for all different flavors including quarks and gluons were not available until recently. In  \cite{Hautmann:2017fcj,Hautmann:2017xtx} a new method for determination of TMD parton densities is described, another method to obtain TMD parton densities from collinear parton densities has been proposed in \cite{Martin:2009ii}, which was applied in \cite{Bury:2017jxo}. In order to fully account for the potential of a TMD parton shower, the initial state kinematics for the hard process calculation should include the transverse momenta. With the development of an automated calculation of multi-leg matrix elements with off-shell initial states \cite{vanHameren:2016kkz} the full potential of TMD parton densities and parton showers can be explored.
Here I will describe application of TMD parton densities, TMD parton showers \cite{Martin:2009ii} \footnote{Very recently new very successful approach to obtain TMD pdfs called Parton Branching method has been developed \cite{Martinez:2018jxt}.} and off-shell matrix elements obtained from \katie\  \cite{vanHameren:2016kkz} to dijets calculations. The advantage of the approach is that the kinematics of matrix elements is not affected. We illustrate the advantage of using TMD densities with off-shell matrix element calculations in an application to azimuthal de-correlations of high \pt\  dijet measurements at the LHC.
The hard matrix elements are calculated as the summed squares of helicity amplitudes, defined following the approach of~\cite{vanHameren:2012if,vanHameren:2013csa} which guarantees gauge invariance.
The parton shower, which is described here, follows consistently the parton evolution of the TMDs. 
By this it is meant that the splitting functions $P_{ab}$, the order in \as , the scale in the calculation of \as\, as well as the kinematic restrictions applied are identical in both the parton shower and the evolution of the parton densities.
A backward evolution method, as now common in Monte Carlo event generators, is applied for the initial state parton shower, evolving from the large scale of the matrix-element process backwards down to the scale of the incoming hadron. However, in contrast to the conventional parton shower, which generates a transverse momentum of the initial state partons during the backward evolution, the transverse momentum of the initial partons of the hard scattering process is fixed by the TMD and the parton shower does not change the kinematics.
The transverse momenta during the cascade follow the behavior of the TMD.  The hard scattering process is obtained directly using off-shell matrix element calculations as described in section. The partonic configuration is stored in the form of an LHE (Les Houches Event) text file, but now including the transverse momenta of the incoming partons. This LHE files are input to the shower and hadronization interface of \cascade (new version \verb+2.4.X+) for the TMD shower where events in HEPMC~ format are produced. \begin{figure}[t!]
\begin{center}
\includegraphics[width=0.45\textwidth]{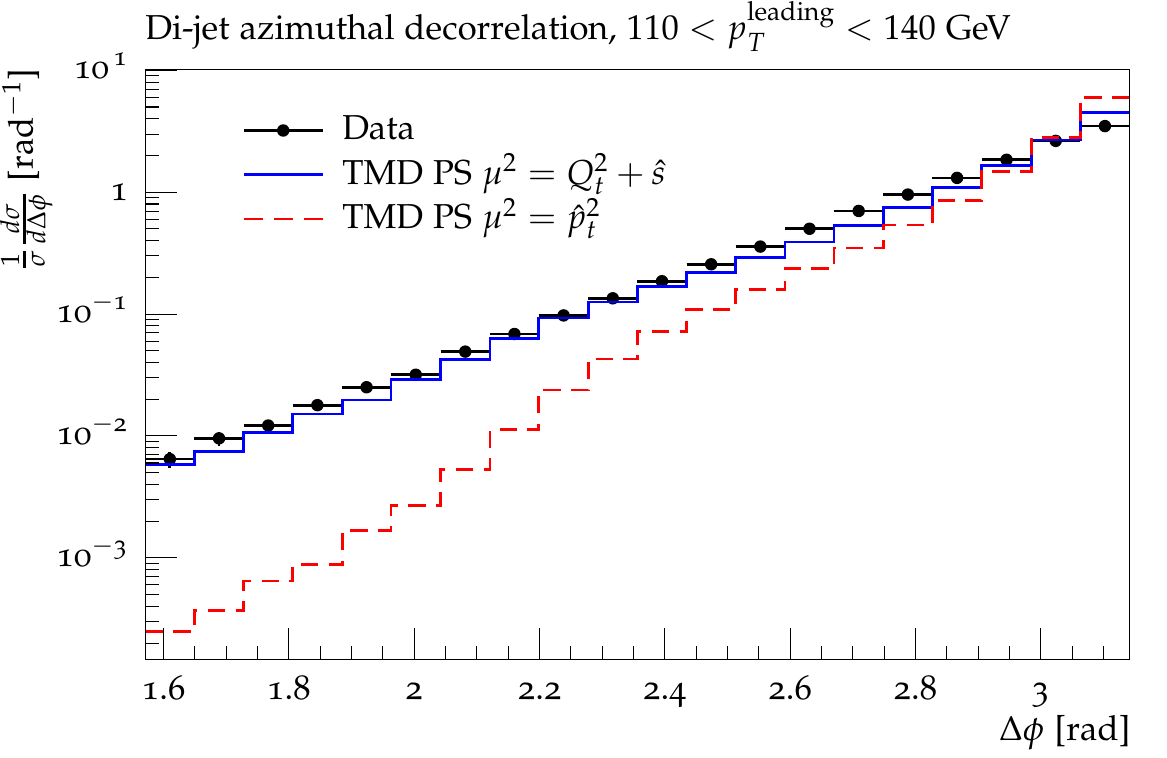} \hskip 1cm
\includegraphics[width=0.45\textwidth]{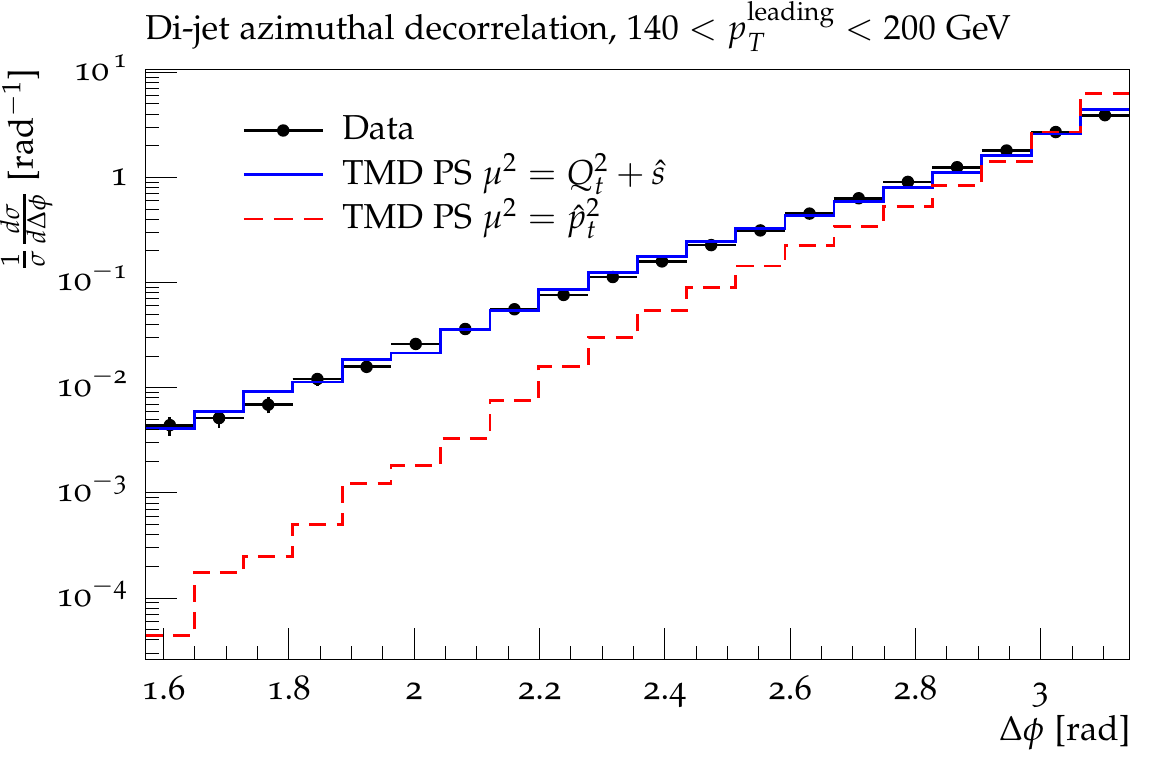}  
\caption{$\Delta \phi$ distribution as measured by \protect\cite{Khachatryan:2011zj} for different regions of $\pt^{leading}$. The data are compared with predictions using off-shell $2\to2$ matrix elements with TMD parton densities, an initial state TMD parton shower, conventional final state parton shower and hadronization. Shown are predictions for two different choices of the factorization scale, as discussed in the text.}
\label{dijets_shower_data}
\end{center}
\end{figure}
Fig.~\ref{dijets_shower_data} shows predictions for the azimuthal de-correlation $\Delta \phi$ for high \pt\ dijets for different regions of $\pt^{leading}$ using TMD parton densities with off-shell matrix elements, parton shower and hadronization in comparison with measurements at $\sqrt{s} = 7$ TeV in pp collisions at the LHC \cite{Khachatryan:2011zj}.  Two different factorization scales are used:
$\mu^2 = Q_t^2 + \hat{s}$, where $Q_t$ is the vectorial sum of the initial state transverse momenta and $\sqrt{\hat{s}}$ is the invariant mass of the partonic subsystem and $\mu^2 =\hat{p_t}^2$.
The first scale choice is motivated by angular ordering (see Ref. \cite{Jung:2003wu}), the second one is the conventional scale choice. The scale choice motivated from angular ordering describes the measurements significantly better than the conventional one. 
It is important to note, that there are no free parameters left: once the TMD parton density is determined, the initial state parton shower follows exactly the TMD parton distribution. The TMD parton distribution is the essential ingredient in the present calculation, and a precise
determination of the TMD parton distribution over a wide range in $x$, \kt\ and scale $\mu$ is an important topic \cite{Hautmann:2017fcj,Hautmann:2017xtx}.
\section{Non-Gaussian broadening in jet quenching}
So far I discussed jets produced in the vacuum. However, jets are copiously produced in heavy ion collisions. Due to their high energies, jets are mostly created in the initial stages of the collisions and travel through a medium such as a Quark Gluon Plasma (QGP).
In particular, quark and gluon jets interact with the strongly interacting medium particles of a QGP. 
Thus, they serve as interesting probes of this kind of medium.
Experimentally, jets are studied via observables on individual jets, such as the nuclear modification factor $R_{AA}$ of jets, as well as via observables on jet pairs.
These dijet observables also allow to study medium effects via the deviations from the momentum balance in the hard collisions. 
Specifically, in \cite{vanHameren:2019xtr} it has been proposed to combine the $k_T$ factorization with the formalism for jet quenching. This allowed for studies of the effects of transverse momentum as generated in initial state on angular decorrelations. 
To be more precise it is necessary to consider that also the partons inside the colliding nucleons have non-vanishing momentum components transverse to the beam axis,
\begin{align}
k_1=x_1\,P_1+k_{1T},&& k_2=x_2\,P_2+k_{2T}\,. 
\end{align}
where $k_1$ and $k_2$ the momenta of the partons inside the colliding nucleons with momenta $P_1$ and $P_2$, and $k_{1T}$ and $k_{2T}$ are the transverse momenta in the laboratory frame and $x_1$ and $x_2$ the momentum fractions. 
Thus, we use unintegrated parton densities rather than parton distribution functions  in order to describe the partons within the colliding nucleons this leads to generalization of the $k_T$ factorization formula.
The medium effect on individual jet particles can be described with a fragmentation function $D(\tilde{x},\mathbf{l},t)$.
For the evolution of the fragmentation function in the medium, we used an equation found by Blaizot, Dominguez, Iancu, and Mehtar-Tani (BDIM)\footnote{For recent developments accounting for shortcomings of the BDIM formalism we refer the reader to \cite{Zakharov:2018rst,Zakharov:2019fro,Zakharov:2019fov,Zakharov:2020sfx,Andres:2020vxs}}
~\cite{Blaizot:2013vha}, which has been solved in \cite{Kutak:2018dim,Blanco:2020uzy}
\begin{equation}
\begin{aligned} 
\frac{\partial}{\partial t} D(\tilde{x},\mathbf{l},t) = & \: \frac{1}{t^*} \int_0^1 dz\, {\cal K}(z) \left[\frac{1}{z^2}\sqrt{\frac{z}{\tilde{x}}}\, D\left(\frac{\tilde{x}}{z},\frac{\mathbf{l}}{z},t\right)\theta(z-\tilde{x}) 
- \frac{z}{\sqrt{\tilde{x}}}\, D(\tilde{x},\mathbf{l},t) \right] \\
+& \int \frac{d^2\mathbf{q}}{(2\pi)^2} \,C(\mathbf{q})\, D(\tilde{x},\mathbf{l}-\mathbf{q},t),
\end{aligned}
\label{eq:ktee1}
\end{equation}
\begin{figure}[t!]
\centering
\includegraphics[width=12.5cm]{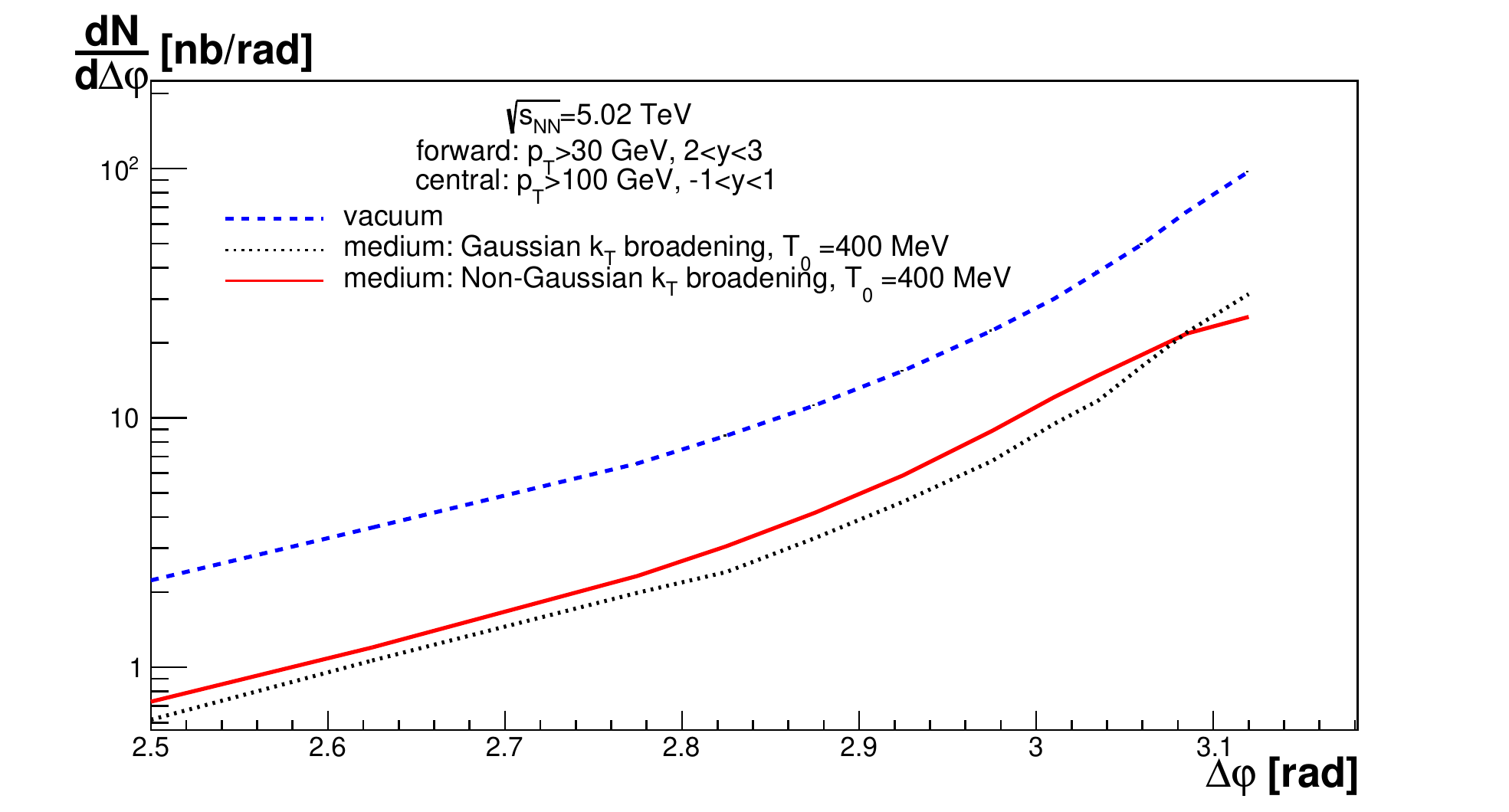}\\
\caption{Upper plot: Azimuthal angular decorrelations between two jets of forward and central rapidities without medium modifications (dashed line), with Gaussian $k_T$ broadening (dotted line) and Non-Gaussian $k_T$ broadening (solid line). Lower plot: Same as the upper one, but normalized to the maximum of the distribution}
\label{Fig:dphi}
\end{figure}
which contains both, a medium induced splitting kernel ${\cal K}$ and a scattering kernel $C$, which describes the transverse momentum transfered by the medium. Where ${\cal K}(z)$ are splitting and  collision kernel $C$ respectively and $\hat{q}$ is the quenching parameter of the medium, $N_c$ is the number of colors and $\alpha_s$ is the QCD coupling constant.
With the above splitting kernel $\cal K$ and the stopping time $t^*$, Eq.~(\ref{eq:ktee1}) describes the medium induced radiations as coherent emissions that take into account multiple scatterings with medium particles which may occur simultaneous to the emission, and their resulting interference effects (cf.~\cite{Blaizot:2012fh}).
Thus, Eq.~(\ref{eq:ktee1}) shows the same kind of suppression for the emission of highly energetic gluons due to interferences as the approach by Baier Dokshitzer Mueller Peign\' e, Schiff, and Zakharov (BDMPS-Z)~\cite{Baier:1996sk,Zakharov:1996fv,Zakharov:1997uu} .
In \cite{vanHameren:2019xtr}, jet-evolution following Eq.~(\ref{eq:ktee1}) was implemented in the Monte-Carlo program \mincas~\cite{Kutak:2018dim}. 
Jet cross section including evolution in medium has been calculated in combined in the medium following the above equation can be also calculated numerically via \mincas.
So far, the BDIM equation, Eq.~(\ref{eq:ktee1}), only exists for gluons, which is why only the production of pairs of gluon jets was addressed. 
Medium was simulated by parametrizing temperature as a function of time. 
For simplicity the temperature dependence that follows from the Bjorken model was used. 
Numerical results for azimuthal decorrelations of di-jets produced in nuclear collisions at $\sqrt{s_{NN}}=5.02$~TeV with one jet going into a forward and the other one in a central rapidity direction ($2<y<3$ with $p_t>30$~GeV and $-1<y<1$ with $p_t>100$~GeV, respectively) are shown in the upper panel of Fig.~\ref{Fig:dphi}. 
As it can be seen the medium effects lead to a considerable suppression of the observed jet-pairs, by at least a factor of $3$ in the region $2.5<\Delta\phi<\pi$ for both results from both cases, Gaussian and the non-Gausian $k_T$ broadening.
Furthermore, the distribution for non-Gaussian $k_T$ broadening appears to be slightly broader than the distribution for Gaussian $k_T$ broadening.
The different widths of the distributions can be seen more easily, when the distributions are normalized to their respective maximum, as it is shown in the lower panel of Fig.~\ref{Fig:dphi}.
As it can be seen, the distributions for Gaussian $k_T$ broadening and the vacuum case show a very similar behavior, while the distribution for the Non-Gaussian $k_T$ broadening is considerably broader.

\section*{Acknowledgements}
I would like to thank the organizers of the Corfu conference for the kind invitation and especially Jan Kalinowski.
The research is founded by Polish National Science Centre grant no. DEC-2017/27/B/ST2/01985.

\bibliographystyle{JHEP}
\bibliography{references}

\end{document}